\documentstyle[12pt]{article}
\begin{document}
%%%%%%%%%%%%%%%%
\begin{titlepage}
\rightline{\vbox{\halign{&#\hfil\cr
&SWAT/45\cr
&\today\cr}}}
\vspace{0.5in}
\begin{center}
{\Large\bf
Duals of nonabelian gauge theories in $D$ dimensions}\\
\medskip
\vskip0.5in

\normalsize {I.G. Halliday and P. Suranyi\footnote{
Permanent address:
Department of
Physics, University of Cincinnati, Cincinnati, Ohio, 45221
USA}
\smallskip
\medskip

{ \sl Department of Physics, University of Wales  Swansea\\
Swansea, Wales SA2 8PP, United Kingdom\\}}

\smallskip
\end{center}
\vskip1.0in

\begin{abstract}  The dual of
an arbitrary $D$-dimensional nonabelian lattice
gauge theory, obtained after character expansion and integration over
the gauge group,   is shown to be a
{\em local} lattice theory in the eigenspace
of the Casimir operators.
For $D\leq4$ we also provide the explicit form of the action as a
product of character expansion coefficients and Racah coefficients.
The representation can be used to facilitate
strong coupling expansions. Furthermore, the possibility of
simulations, at weak coupling, in the dual representation,
 is also discussed.
\end{abstract}
\end{titlepage}
 Ever since their invention  by
Wilson~\cite{wilson},
lattice gauge theories have inspired research in a multitude of
directions. One of the most interesting of these has been the
investigation
of duality properties.   The study of duality properties of various
spin models has a long history, but more recently duals of
compact $U(1)$
 lattice gauge theories were also found~\cite{banks}~\cite{savit}.

The first step of finding duals of lattice gauge theories is
always the character expansion of the Wilson-action. The second
step is integration over the gauge degrees of freedom. In the
case of $U(1)$ theories these steps lead to integer valued fields
satisfying constraints. The constraints have the form of
linear equations for the eigenvalues of Casimir operators. For
$U(1)$ theories these are additive quantum numbers.

Suppose now
that the dual picture is to be used for simulations. Then
either the additive quantum numbers involved in the constraints
have to be summed over, or the constraints (Bianchi identities)
have to be resolved in terms of fields with no
constraints. The first of these routes is not feasible for
$U(1)$ gauge theories: it would be tantamount to the exact solution
 of the theory. However, the second route has been
successful~\cite{banks}~\cite{savit}.

When a similar procedure is applied to nonabelian  lattice gauge
theories the above two steps also lead to local discrete valued
models. The partition function of such a model is formed from a sum
over products of the character expansion coefficients, depending on
the eigenvalue of the
quadratic Casimir operator and of Clebsch-Gordan coefficients
depending on two kinds of discrete valued fields. Fields belonging
to the first group take values that are the eigenvalues of diagonal
elements of the
Lie algebra, labeling states in a given representation. We will
denote the collection of them for a particular group element by
$m,n,M,N,\mu$, or $\nu$.
These are additive quantum numbers and
they all satisfy constraints in the form of linear equations.
Either these constraints have to be resolved, or the summation
over the corresponding fields has to be performed. The second
group of discrete valued fields take values that are the
eigenvalues of Casimir
operators. They will be labeled by letters $j,J,$ and $l$.
They satisfy triangular and tetrahedral inequalities.
As will be pointed out later, such constraints would allow
simulations respecting detailed balance. For short we
will refer to these fields as magnetic quantum number and
angular momentum, respectively, though these names are truly
appropriate only if the gauge group is $SU(2)$.

The purpose of this letter is to investigate the possibility of
performing summations over magnetic quantum numbers. Of course,
this can be done in a formal manner, but the crucial question is
whether the resulting effective theory of angular momentum valued fields is
local or not. We will answer this question in the affirmative
for arbitrary
compact gauge group and arbitrary space-time dimension D. In fact,
the statement
is fairly obvious for the more or less trivial case $D=2$, and
it has recently been proved for $D=3$ by several
authors~\cite{boulatov}~\cite{indian}.

A lattice gauge theory in the fundamental representation (a
generalization to other representations is possible) has
the form
\begin{equation}
Z=\prod_{\bf r}\left[\prod_{i=1}^D\int dA_i({\bf r})
\prod_{i>k\geq1}^D\exp\{ \beta\chi_f[(U_{ik}({\bf r})]\}\right],
\label{part}
\end{equation}
where $\chi_f$ denotes the character of the group in
the fundamental representation depending on the plaquette
element, to be described later, and $\int dA_i({\bf r})$ represents
integration over the Haar measure. The product index $i$ represents
the axes of the lattice, while the $(ik)$ pair, in conjunction
with ${\bf r}$, represents a
particular plaquette (2 dimensional simplex) of the lattice.

The exponentials can be expanded in a series of characters to
give
\begin{equation}
\exp\{ \beta\chi_f[(U_{ik}({\bf r})]\}=\sum_j d_j\chi_j[U_{ik}
({\bf r})]
c_j(\beta)
\label{char}
\end{equation}
where $d_j$ is the dimension of the representation labeled
by $j$.

Now the characters are traces of the representation matrix of
the plaquette element. The plaquette element can be written
as the product of the link elements at the boundary of the
plaquette. The link element $U_i({\bf r})$ is
associated with a link originating at ${\bf r}$
and extending one lattice unit into the direction of the $i$th
axis. The element  $U_i({\bf r})$ is also
associated with the positive direction, while its adjoint
with the negative direction. Setting ${\bf r}=0$, for
simplicity, we have
\begin{equation}
\chi_j[U_{ik}
(0)]={\rm Tr}_j[U_i^\dagger(0)U_k(0)U_i({\bf \hat e}_k)
U_k^\dagger({\bf \hat e}_i)],
\label{plaq}
\end{equation}
where ${\bf \hat e}_k$ is a unit lattice vector in the positive
direction
along the $k$th axis.

The dual of the lattice consists of $D$ dimensional
hypercubes centered at the vertices of the original lattice.
The boundary of a hypercube
consists of $2D$ codimension 1 cubes. The dual of a link of
the original lattice is such a cube, on which the link variables
$U_i$ live. The duals of plaquettes are codimension 2 simplexes (for
$D$=4, plaquettes themselves). They are the boundaries of the
$D-1$ dimensional cubes that themselves
form the boundary of the hypercube.
Duals of plaquettes attached to a single link of the original lattice
form the bondary of such a $D-1$ dimensional simplex.
There are $2D(D-1)$ such simplexes attached to each hypercube.

Our discussion starts with a simple observation:

\begin{quote} {\it 1) Every pair of adjacent link variables of
a plaquette define a unique hypercube. The four possible pairs
one can form on a loop define four adjacent hypercubes.}
\end{quote}

The proof is simple. Two adjacent links of the original
lattice intersect in a lattice point. Let us associate this point
with the pair of adjacent link variables.
A point of the original lattice
is, however,
 dual to a hypercube (centered at the point). The corners
of a plaquette define four separate, but adjacent hypercubes.
Their mutual boundaries are the duals of the link, codimension
1 simplexes.

Now saturate each product of link matrices on the plaquette by
a complete system of states. Each complete system can be labeled
by $(j,m)$, where $m$ represents the collection of
quantum numbers labeling states in representation $j$. Then
it follows that
\begin{quote} {\it 2) Every magnetic quantum number,
$m$ is  associated with a unique hypercube.}
\end{quote}
This is true because every $m$ is
associated with a pair of link matrices, but according
to 1) every pair of adjacent link matrices is associated
 with a hypercube.

The plaquette
contribution can then be written as
\begin{equation}
\chi_j[U_{ik}
(0)]=D^j_{m_1m_2}(U_i^\dagger)D^j_{m_2m_3}(U_k)
D^j_{m_3m_43}(U_i)
D^j_{m_4m_1}(U_k^\dagger),
\label{plaq2}
\end{equation}
where the arguments of the link operators were dropped for
simplicity. They are the same as in (\ref{plaq}). Each of the
four indices $m_i$ is associated with an adjacent hypercube,
according to 1) and 2).

Fig. 1. shows a plaquette in the $ik$ plane. It also shows the
magnetic quantum numbers labeling systems saturating
adjacent products of $U$-operators. The magnetic quantum numbers
are associated with dual hypercubes centered at corners of the
plaquette.

It is worth making a comment at this point. Lemma 1 fails
 for globally symmetric nonabelian spin
models. There is no natural
way of associating the two hypercubes, connected by a link, with
one or the other of the set of variables $m_1$ and $m_2$
of the product
$D^j_{m_1m_2}(U^\dagger)D^j_{m_2m_1}(U') $. Indeed,
their duals are not local in the space of Casimir operators. The
results of this paper would fail for nonabelian spin models.

Furthermore, it is easy to see that
\begin{quote} {\it 3) Every link matrix $U_i$ is associated
with the same hypercube on its left (and also on its right) in
every plaquette it enters.}
\end{quote}

The proof follows from the definition of link variables.  Naturally,
the association is reversed for the matrix $U_i^\dagger$.
Then we obtain
\begin{quote} {\it 4) For every link variable, $U$, the
left subscript (right subscript) of each rotation function
$D^j_{m'm}(U)$ it enters is associated with the same hypercube.}
\end{quote}

4) is a simple consequence of 2) and 3).

Let us prove now the most important result of this letter:
\begin{quote} {\it 5) Every Clebsch-Gordan coefficient,
obtained after combining rotation functions and
integrating over the gauge variables,
depends on magnetic quantum numbers associated with a
single dual hypercube only. The angular momentum variables  of
the Clebsch-Gordan coefficients are shared by neighboring
hypercubes only.}
\end{quote}

The proof starts with the
 use of the addition theorem for representation functions
to reduce the number of representation functions $(2D-2)$ depending on
the particular link $U$.
\begin{equation}
D^{j_1}_{m_1n_1}(U)D^{j_2}_{m_2n_2}(U)=\sum_{J}
(2J+1)(-1)^{2(J+j_2-j_1)}\left(\begin{array}{ccc} j_1&j_2&M\\
m_1&m_2&J\end{array}
\right)D^{J}_{MN}(U)
\left(\begin{array}{ccc} j_1&j_2&N\\						n_1&n_2&J\end{array}
\right),
\label{compose}
\end{equation}
where the Wigner's three-$j$ symbols have been used  rather
then Clebsch-Gordan coefficients.

When a rotation function  with
argument $U$ is combined with another one with argument $U^\dagger$,
then the relation
\begin{equation}
D^{j}_{mn}(U^\dagger)=(-1)^{m-n}D^j_{-n-m}(U)
\label{conj}
\end{equation}
should be first used.

Two important observations concerning
(\ref{compose}) should be made at
this point. First, the dependence on indices $m$ and $n$,
associated with different hypercubes, factorizes in the two
three-$j$ symbols. Second, the new rotation
function appearing in the addition theorem also satisfies
4), because $M=m_1+m_2$ and $N=n_1+n_2$.

Subsequent applications of the addition theorem reduce
the product of $2(D-1)$ rotation functions to a single
rotation function $D^J_{\mu\nu}(U)$ and a product of
2$(2D-3)$ three-$j$ symbols, $2D-3$ of which
depend on 2$(D-1)$  the quantum numbers $m_i$ only,
which are associated with one of the
hypercube and the same number of them depend on quantum
numbers $n_i$ only, which are associated with the other hypercube.
Finally, integration over the group implies that $J=\mu=\nu=0$
in the last rotation function. This, in turn, implies that
the last two three-$j$ symbols that contain the
quantum numbers $(J,\mu)$ and $(J,\nu)$, respectively,
  turn into Kronecker deltas for two pairs  of
angular momentum vectors.
That leaves us with the product of $2(D-2)$ three-$j$ symbols
dependent on quantum numbers $m_i$ only and the same number of symbols
dependent on the quantum numbers $n_i$ only. The dependence
on additive quantum numbers labeling states is completely
factorized.

Notice now that a similar construction can be performed on every
one of the $2D$ links emanating from a given lattice point.
Altogether, we will have 4$D(D-2)$ three-$j$ symbols
dependent on $m_i$ quantum numbers associated with a given
hypercube only. Furthermore, the angular momenta in these three-$j$
symbols are shared by the two hypercubes joining in the $D-1$
dimensional simplex, dual to the appropriate link variable, only.
This completes the proof of the theorem.

Now the final form of the main result of the paper can be spelled
out as
\begin{quote}
{\it 6) Summation over magnetic quantum numbers results in a discrete
local field theory in the eigenvalues of Casimir operators.}
\end{quote}

Summation over the magnetic quantum numbers results in
factors associated with  hypercubes. The factor depends on angular
momenta corresponding to the $2D$
codimension 1 simplexes of the hypercube and additional 2$D(D-2)$
angular momenta obtained
at repeated applications of addition theorem (\ref{compose}),
also associated with the same hypercube. Thus, the range of
the interactions is the size of a single hypercube.

Let us investigate now the simplest cases $D=2,3$, and 4.
Although $D=2$ lattice gauge theories are trivial, still the
application of our results to them is instructive. Also
the algebra is simple enough so that it can
 be written out in detail. There are $D=2$ link variables
$U_1$ and $U_2$ running out of an arbitrary point of the original
lattice and $D=2$ variables
$U_{\bar1}$ and $U_{\bar2}$ running into it. The $2D(D-1)=4$
plaquettes containing these variables are
\begin{equation}
{\rm Tr}(U_{\bar2}U_1...),~~{\rm Tr}(U_1^\dagger U_2...),~~
{\rm Tr}(U_2^\dagger U_{\bar1}^\dagger...),~~
{\rm Tr}( U_{\bar1}U_{\bar2}^\dagger...),
\label{plaquettes}
\end{equation}
where link operators connecting other dual
plaquettes (hypercubes) have been omitted from the traces.

After inserting complete systems of states one obtains
the following product of rotation functions:
\begin{eqnarray}
&\sum_{m_{1\bar2}m_{12}m_{\bar12}m_{\bar1\bar2}
}&D^{j_{1\bar2}}_{...m_{1\bar2}}(U_{\bar2})
D^{j_{1\bar2}}_{m_{1\bar2}...}(U_1)
D^{j_{12}}_{...m_{12}}(U_{1}^\dagger)
D^{j_{12}}_{m_{12}...}(U_2)\nonumber\\
&\times&D^{j_{\bar12}}_{...m_{\bar12}}(U_{2}^\dagger)
D^{j_{\bar12}}_{m_{\bar12}...}(U_{\bar1}^\dagger)
D^{j_{\bar1\bar2}}_{...m_{\bar1\bar2}}(U_{\bar1})
D^{j_{\bar1\bar2}}_{m_{\bar1\bar2}...}(U_{\bar2}^\dagger),\label{dees}
\end{eqnarray}
where indices associated by other dual plaquettes have been omitted.

Integration over the rotation functions is simple, because there are
only two functions for each rotation, each giving a Kronecker delta
for the angular momenta and magnetic quantum numbers as well. Thus,
the summation over magnetic quantum numbers results in a factor
$2j+1$ only, where $j=j_{12}=...=j_{\bar1\bar2}$. The factor
corresponding to a dual plaquette is proportional to a Kronecker
deltas for all the angular momenta involved and a trivial
$j$-dependent multiplier. Since the angular momenta are shared
between neighboring dual plaquettes, the partition function
becomes diagonal in angular momentum. Thus, the partition function is
$\sum_j(2j+1)[c_j(\beta)]^V$, where $V$ is the volume
of the lattice.

The $D=3$ case~\cite{boulatov}~\cite{indian}
 is more complicated. There are $2D=6$ link
(dual plaquette) variables, and $2D(D-1)=12$ plaquettes (dual
links) involved. Each link variable appears in four different
plaquettes. Each appearance contributes to the expression by a rotation
function, as in (\ref{dees}). The four rotation
functions can be combined pairwise using (\ref{compose})
 and (\ref{conj})
giving two three-$j$ symbols for the dual
plaquette in question and two rotation functions. Integration
over the group space results in the identification of the
new angular momenta and magnetic quantum numbers in the
three-$j$ symbols. Before we write down the appropriate
expressions obtained after this procedure we introduce
a concise notation,
due to Wigner~\cite{wigner}. Magnetic quantum numbers
will be omitted from three-$j$ symbols, with the
understanding that repeated angular momenta in products
imply summation over the corresponding magnetic quantum numbers. Since
we restrict our discussion to three-$j$ symbols
associated with a single hypercube, every angular momentum
will be uniquely associated with a  magnetic quantum number.\footnote{
As Wigner points it out~\cite{wigner}
 contravariant and covariant components should be used in the
three-$j$ symbols, which differ in sign.
Contravariant component are always contracted with covariant ones.
Our final result is
not affected by this complication, since
we intend to sum over {\em all} magnetic numbers. Thus, for the purpose
of simplifying notations we will not distinguish contravariant
and covariant components.}

Labeling angular momenta as in (\ref{dees}) we obtain the following
product of three-$j$ symbols:
\begin{eqnarray}
&&(j_{1\bar3}j_{12}J_1)(j_{1\bar2}j_{13}J_1)(j_{12}j_{2\bar3}J_2)
(j_{\bar12}j_{23}J_2)(j_{\bar13}j_{23}J_3)(j_{\bar23}j_{13}J_3)
\nonumber\\&\times&(j_{\bar1\bar2}j_{\bar1\bar3}J_{\bar1})
(j_{\bar13}j_{\bar12}J_{\bar1})(j_{\bar1\bar2}j_{\bar2\bar3}J_{\bar2})
(j_{\bar23}j_{1\bar2}J_{\bar2})(j_{\bar1\bar3}j_{\bar2\bar3}J_{\bar3})
(j_{2\bar3}j_{1\bar3}J_{\bar3}),
\label{threed}
\end{eqnarray}
This is the complete expression for the contribution of a
dual cube, apart from the multiplier $\prod (2J_i+1)$, shared
by neighboring dual cubes.

Using the relations
\begin{equation}
(j_1j_2J_3)(j_2j_3J_1)(j_3j_1J_2)=
\left\{\begin{array}{ccc} J_1&J_2&J_3\\
j_1&j_2&j_3\end{array}
\right\}(J_1J_2J_3)
\label{racah}
\end{equation}
and
\begin{equation}
(j_1j_2J_3)(j_2j_3J_1)(j_3j_1J_2)(J_1J_2J_3)=
\left\{\begin{array}{ccc} J_1&J_2&J_3\\
j_1&j_2&j_3\end{array}
\right\}
\label{racah2}
\end{equation}
(\ref{threed}) reduces to
\begin{equation}
\left\{\begin{array}{ccc} J_{\bar1}&J_2&J_3\\
j_{23}&j_{\bar13}
&j_{\bar12}\end{array}
\right\}\left\{\begin{array}{ccc} J_1&J_{\bar2}&J_3\\
j_{\bar23}&j_{13}&j_{1\bar2}
\end{array}
\right\}\left\{\begin{array}{ccc} J_1&J_2&J_{\bar3}\\
j_{2\bar3}&j_{1\bar3}&j_{12}
\end{array}
\right\}\left\{\begin{array}{ccc} J_{\bar1}&J_{\bar2}&J_{\bar3}\\
J_{1}&J_2&J_3\end{array}
\right\}\left\{\begin{array}{ccc} J_{\bar1}&J_{\bar2}&J_{\bar3}\\
j_{\bar2\bar3}&j_{\bar1\bar3}&j_{\bar1\bar2}\end{array}
\right\}.
\label{final}\end{equation}
(\ref{final}) has a simple geometrical interpretation if
the gauge group is $SU(2)$~\cite{boulatov}
{}~\cite{indian}. Since angular momenta are attached to plaquettes on
the original lattice, and links (vectors)
on the dual lattice, each six-$j$
symbol represents a tetrahedron. The five tetrahedrons
corresponding to the five six-$j$ symbols form a cube-like
object with triangular faces.
The angular momenta $J_i$ and $J_{\bar i}$ correspond to face
diagonals of the dual cube. All angular momenta are shared by
neighboring dual cubes. The angular momenta $j_{ik}$ are
shared by three other, neighboring cubes: the ones displaced by
${\bf\hat e}_i,{\bf\hat e}_k,$ and by ${\bf\hat e}_i
+{\bf\hat e}_k$. Barred indices correspond to negative
directions. Each of the angular momenta $J_i$ is shared only by
one other cube, the one displaced by vector ${\bf\hat e}_i$. The
collection of tetrahedra cannot always be embedded in flat three
dimensional space~\cite{ponzano}. In fact, if the coefficients
$c_j(\beta)$ of (\ref{char}) are omitted then the partition
function reduces to the discrete model of Regge for three dimensional
gravity.~\cite{boulatov}~\cite{indian}

Let us turn now to the physically most interesting $D=4$ case.
Integration over the gauge degrees of freedom leads to the
following invariant sum over products of three-$j$ symbols:
\begin{eqnarray}
&&(j_{12}j_{2\bar4}J_2^{1\bar4})
(j_{12}j_{1\bar4}J_1^{2\bar4})
(j_{1\bar4}j_{2\bar4}J_{\bar4}^{12})
(j_{34}j_{23}J_3^{24})
(j_{24}j_{23}J_2^{34})
(j_{24}j_{34}J_4^{23})
\nonumber\\&\times&(j_{\bar1\bar2}j_{\bar24}J_{\bar2}^{\bar14})
(j_{\bar1\bar2}j_{\bar14}J_{\bar1}^{\bar24})
(j_{\bar14}j_{\bar24}J_4^{\bar1\bar2})
(j_{\bar2\bar4}j_{\bar3\bar4}J_{\bar4}^{\bar2\bar3})
(j_{\bar2\bar3}j_{\bar3\bar4}J_{\bar3}^{\bar2\bar4})
(j_{\bar2\bar3}j_{\bar2\bar4}J_{\bar2}^{\bar3\bar4})
\nonumber\\&\times&(j_{\bar13}j_{3\bar4}J_3^{\bar1\bar4})
(j_{\bar13}j_{\bar1\bar4}J_{\bar1}^{3\bar4})
(j_{\bar1\bar4}j_{3\bar4}J_{\bar4}^{\bar13})
(j_{1\bar3}j_{\bar34}J_{\bar3}^{14})
(j_{1\bar3}j_{14}J_1^{\bar34})
(j_{14}j_{\bar34}J_4^{1\bar3})
\nonumber\\&\times&(j_{1\bar2}j_{\bar23}J_{\bar2}^{13})
(j_{1\bar2}j_{13}J_1^{\bar23})
(j_{13}j_{\bar23}J_3^{1\bar2})
(j_{\bar12}j_{2\bar3}J_2^{\bar1\bar3})
(j_{\bar12}j_{\bar1\bar3}J_{\bar1}^{2\bar3})
(j_{\bar1\bar3}j_{2\bar3}J_{\bar3}^{\bar12})
\nonumber\\&\times&(J_1^{2\bar4}J_1^{\bar34}J_1^{\bar23})
(J_2^{1\bar4}J_2^{34}J_2^{\bar1\bar3})
(J_3^{24}J_3^{\bar1\bar4}J_3^{1\bar2})
(J_4^{\bar1\bar2}J_4^{1\bar3}J_4^{23})
\nonumber\\&\times&(J_{\bar1}^{\bar24}J_{\bar1}^{3\bar4}
J_{\bar1}^{2\bar3})
(J_{\bar2}^{\bar14}J_{\bar2}^{\bar3\bar4}J_{\bar2}^{13})
(J_{\bar3}^{\bar2\bar4}J_{\bar3}^{14}J_{\bar3}^{\bar12})
(J_{\bar4}^{12}J_{\bar4}^{\bar13}J_{\bar4}^{\bar2\bar3}).
\label{fourd}
\end{eqnarray}
Again the symbols $j$ correspond to angular momenta carried
by plaquettes, while symbols $J$ represent angular momenta
obtained when the $j$-s are added. Both of these angular momenta
are shared by neighboring hypercubes, exactly the same manner
as for $D=3$. Note that the subscript of symbols $J$ indicates
the direction of the neighboring hypercube sharing the
corresponding angular momentum.

Using relation (\ref{racah}), (\ref{racah2}), and~\cite{wigner}
\begin{equation}
(j_1j_2j)(j_3j_4j)=(-1)^{2j_3}\sum_{j'}(2j'+1)
\left\{\begin{array}{ccc} j_{1}&j_{4}&j'\\
j_{3}&j_2&j\end{array}
\right\}(j_1j_4j')(j_3j_2j')
\label{racah3}
\end{equation}
one can express product (\ref{fourd}) by means of invariant,
six-$j$ symbols. We obtain
\begin{eqnarray}
&&\sum_{J_1,...,J_7}\prod_{i=1}^7(2J_i+1)
\left\{\begin{array}{ccc} J^{1\bar4}&J^{2\bar4}&J^{12}\\
j_{1\bar4}&j_{2\bar4}&j_{12}\end{array}\right\}					\left\{\begin{array}{ccc}
J^{23}&J^{34}&J^{24}\\
j_{23}&j_{34}&j_{24}\end{array}\right\}
\left\{\begin{array}{ccc} J^{\bar1\bar2}&J^{\bar14}&J^{\bar24}\\
j_{\bar1\bar2}&j_{\bar14}&j_{\bar24}\end{array}\right\}
\nonumber\\ &\times&								\left\{\begin{array}{ccc}
J^{\bar2\bar3}&J^{\bar2\bar4}&J^{\bar3\bar4}\\
j_{\bar2\bar3}&j_{\bar2\bar4}&j_{\bar3\bar4}\end{array}\right\}
\left\{\begin{array}{ccc} J^{\bar13}&J^{\bar1\bar4}&J^{3\bar4}\\
j_{\bar13}&j_{\bar1\bar4}&j_{3\bar4}\end{array}\right\}
\left\{\begin{array}{ccc} J^{1\bar3}&J^{14}&J^{\bar34}\\
j_{1\bar3}&j_{14}&j_{\bar34}\end{array}\right\}
\left\{\begin{array}{ccc} J^{13}&J^{1\bar2}&J^{\bar23}\\
j_{13}&j_{1\bar2}&j_{\bar23}\end{array}\right\}
\nonumber\\&\times&
\left\{\begin{array}{ccc} J^{\bar12}&J^{\bar1\bar3}&J^{2\bar3}\\
j_{\bar12}&j_{\bar1\bar3}&j_{2\bar3}\end{array}\right\}
\left\{\begin{array}{ccc} J^{3\bar4}&J^{1\bar3}&J_1\\
J^{\bar12}&J^{\bar24}&J^{2\bar3}\end{array}\right\}
\left\{\begin{array}{ccc} J^{\bar14}&J^{23}&J_2\\
J^{1\bar3}&J^{\bar24}&J^{\bar1\bar2}\end{array}\right\}
\left\{\begin{array}{ccc} J^{\bar2\bar4}&J^{\bar34}&J_3\\
J^{1\bar3}&J^{\bar12}&J^{14}\end{array}\right\}
\nonumber\\&\times&
\left\{\begin{array}{ccc} J_1&J_2&J_3\\
J^{1\bar3}&J^{\bar12}&J^{\bar24}\end{array}\right\}
\left\{\begin{array}{ccc} J^{3\bar4}&J^{1\bar4}&J_4\\
J^{34}&J_1&J^{\bar1\bar3}\end{array}\right\}
\left\{\begin{array}{ccc} J_3&J^{\bar14}&J_5\\
J^{23}&J_1&J_2\end{array}\right\}
\left\{\begin{array}{ccc} J_4&J_5&J^{24}\\
J^{23}&J^{34}&J_1\end{array}\right\}
\nonumber\\&\times&
\left\{\begin{array}{ccc} J^{\bar13}&J^{1\bar4}&J_6\\
J_4&J^{\bar1\bar4}&J^{\bar34}\end{array}\right\}
\left\{\begin{array}{ccc} J_5&J_6&J^{1\bar2}\\
J^{\bar1\bar4}&J^{24}&J_4\end{array}\right\}
\left\{\begin{array}{ccc} J^{2\bar4}&J_6&J^{\bar2\bar3}\\
J^{\bar13}&J^{12}&J^{1\bar4}\end{array}\right\}
\left\{\begin{array}{ccc} J^{\bar2\bar3}&J^{\bar34}&J_7\\
J^{\bar23}&J_6&J^{\bar24}\end{array}\right\}
\nonumber\\&\times&
\left\{\begin{array}{ccc} J_7&J^{13}&J_5\\
J^{1\bar2}&J_6&J^{\bar23}\end{array}\right\}
\left\{\begin{array}{ccc} J_7&J_3&J^{\bar3\bar4}\\
J^{\bar14}&J^{13}&J_5\end{array}\right\}
\left\{\begin{array}{ccc} J_7&J_3&J^{\bar3\bar4}\\
J^{\bar2\bar4}&J^{\bar2\bar3}&J^{\bar34}\end{array}\right\},
\label{fourdfinal}
\end{eqnarray}
	where the subscripts of the $J^{ab}_i$ angular momenta,
appearing in
(\ref{fourd}) have been omitted. The angular momenta $J_1$,...,$J_7$
appear in the process of reduction of (\ref{fourd}). They are
not shared by neighboring hypercubes.

Finally, we examine the possibility of simulations, using the
dual representation. In principle, such a possibility
is exciting, because simulating integer valued theories
without the need of tedious matrix multiplications makes
computations faster. There are several issues, however,
which have to be resolved before simulations can be attempted.

The first issue is detailed balance. Suppose an angular momentum
$j$ is updated. Suppose $\Delta$ is the range of the allowed
random change, i.e. the randomly proposed new value of
$j$ satisfies $j-\Delta\leq j'\leq j+\Delta$ that is
{\em independent of the inequalities that $j$ satisfies}. Then there are
two possibilities. Either $j'$ also satisfies all the required
inequalities, or it does not. In the former case the probability
of the $j\rightarrow j'$ transition is the same as that of
the transition $j'\rightarrow j$. In the latter case the two
probabilities are still equal, namely they are zero. Thus for
such an update procedure the detailed balance condition is satisfied.

The second issue is the complicated form of
six-$j$ symbols. Note, however, that the
 aim is to perform simulations in the weak
coupling regime, relevant in the continuum limit. At weak
coupling the average angular momentum is large. In fact,
$j\sim\sqrt{\beta}$. For large values of the angular momentum
the semiclassical limit of six-$j$ symbols can be used.
Wigner~\cite{wigner} has shown that in that limit six-$j$ symbols
are given in an average sense by
\begin{equation}
\left\{\begin{array}{ccc} j_1&j_2&j_3\\
J_1&J_2&J_3\end{array}
\right\}^2\simeq\frac{1}{4\pi|({\bf j}_1\times {\bf j_2})
\cdot{\bf J_3}|},
\label{sixj}
\end{equation}
i.e by the inverse of 24$\pi$ times the volume, $V$, of the
tetrahedron. The volume can be expressed by the edges of
the tetrahedron using Cayley's formula
\begin{equation}
288V^2=\left|\begin{array}{ccccc}
			0&J_1^2&J_2^2&J_3^2&1\\
			J_1^2&0&j_3^2&j_2^2&1\\
			J_2^2&j_3^2&0&j_1^2&1\\
    J_3^2&j_2^2&j_1^2&0&1\\
1&1&1&1&0\end{array}\right|.
\label{volume}
\end{equation}
The volume has square root type zeros at the edge of the
allowed region,
thus, the six-$j$ symbols have mild, integrable singularities
as the tetrahedron becomes degenerate.

One may also consider using continuous angular momentum variables.
This would be approximately equivalent to using a
noncompact gauge group. It should not alter results in
the weak coupling limit.

One very important issue we have not yet resolved is the positivity
of the integrand of the functional integral in
$j$-representation. Indeed, (\ref{sixj})
gives an expression for the square of six-$j$ symbols only.
The
six-$j$ symbols themselves oscillate.
The phase of oscillation was found by Ponzano and Regge~\cite{ponzano}
{}~\cite{bidernharn}.  In the semiclassical limit,
summations over angular momenta are dominated by values
which make all phases of oscillations stationary.
The form of the resulting field theory
is complicated (at least in the $D=4$ case) and will be dealt with
in a future publication.

\section*{ACKNOWLEDGEMENTS}

The authors thank A. Schwimmer for valuable discussions.
The work of P. S. has been
supported by funds from the
Research Committee of the
Higher Education Funding Council for Wales (HEFCWR).
  He is also indebted for the partial support of
the United States Department of Energy under
grant no.  DE-FG02-84ER40153.

\newpage
\begin{center}{\bf \large Figure Captions}\end{center}
Fig.1. A plaquette in the $ik$ plane. $m_i$ represent magnetic
quantum numbers associated with dual hypercubes.

\begin{thebibliography}{10}
\small
\addtolength{\itemsep}{-6pt}
\bibitem{wilson} K. G. Wilson,  {\sl Phys. Rev.} {\bf D14} (1974)
2455;
\bibitem{banks} T. Banks, J. Kogut, and R. Myerson, {\sl Nucl. Phys.}
{\bf B129} (1977) 493;
\bibitem{savit} R. Savit, {\sl Rev.Mod.Phys.} {\bf 52} (1980) 453;
\bibitem{boulatov} D.V. Boulatov, {\sl Int. Journ. Mod. Phys.}
{\bf A8} (1993) 3139
\bibitem{indian} R. Anishetty, S.Cheluvaraja, H.S.Sharatchandra,
and M. Mathur, {\sl Phys. Lett.} {\bf B314} (1993) 387
\bibitem{wigner} E.P. Wigner:{\sl  Group Theory}, Academic Press, New York
1959.
\bibitem{ponzano} G. Ponzano and T. Regge, ``Semiclassical
limit of Racah coefficients,'' in {\sl Spectroscopic and Group
Theoretical Methods in Physics} (Racah Memorial Volume), F. Bloch
et. al. editors, North Holland, Amsterdam, 1968;
\bibitem{bidernharn} L. C. Biedernharn and J. D. Louck, ``{\sl The
Racah-Wigner Algebra in Quantum Theory.}'' (Encyclopedia of
Mathematics, Vol. 9) Adddison-Wesley, Reading, Massachusetts, 1981;
\end{thebibliography}
\end{document}